\documentclass[sort&compress,preprint,12pt]{elsarticle}

\usepackage{amssymb}
\usepackage{amsmath}

\journal{Environmental International}

\begin{document}

\begin{frontmatter}



\title{Advancing Atmospheric Pollution Monitoring with Airborne THz Spectrometer}

\author[1]{Candida Moffa\corref{cor1}}
\ead{candida.moffa@uniroma1.it}
\cortext[cor1]{Corresponding authors}
\author[1]{Alessandro Curcio}
\author[1]{Camilla Merola}
\author[1,2]{Daniele Francescone}
\author[1]{Marco Magi}
\author[1]{Massimiliano Coppola}
\author[1,2]{Lucia Giuliano}
\author[1,2]{Mauro Migliorati}
\author[3]{Massimo Reverberi}
\author[1]{Leonardo Mattiello}
\author[1,2]{Massimo Petrarca \corref{cor1}}
\ead{massimo.petrarca@uniroma1.it}
\affiliation[1]{organization=Department of Basic and Applied Sciences for Engineering (SBAI), Sapienza, University of Rome
            addressline={Via Antonio Scarpa, 16}, 
            city={Rome},
            postcode={00161}, 
            state={Italy},
            country={}
}
\affiliation[2]{organization=Roma1-INFN,
addressline={Piazzale Aldo Moro, 2}, 
city={Rome},
postcode={00185}, 
state={Italy},
country={}
}
\affiliation[3]{organization=Department of Environmental Biology (DBA), Sapienza, University of Rome
            addressline={Piazzale Aldo Moro, 5}, 
            city={Rome},
            postcode={00185}, 
            state={Italy},
            country={}
}
%% Abstract
\begin{abstract}
%In this work, we present a THz-on-drone-based detector of air pollutants capable of real-time remote distance measurements at different altitudes. The system combines a highly stable unmanned aerial system (UAS, drone) and THz-CW laser technology, allowing for flexible, high-resolution remote spectroscopy measurements at the point of care.
%The system presents high stability and is not affected by the UAS’s vibrations in flight notwithstanding the altitude from the soil allowing for the detection of multiple substances. The presented proof-of-concept system demonstrates that it is possible to employ the combined UAS-THz-CW for environmental pollutants detection in the atmosphere.

This study details the development and validation of an airborne THz spectrometer designed for real-time, remote detection of atmospheric pollutants. The platform couples a stabilized unmanned aerial system (UAS) with a high-resolution terahertz continuous-wave (THz-CW) laser source and detector, enabling flexible, in situ spectroscopic analysis of dispersed atmospheric vapours. This proof-of-concept investigation confirms the feasibility of UAS-THz-CW systems for spatially resolved environmental pollutant monitoring. The demonstrated capability of this terahertz sensor for remote, multi-component detection of atmospheric contaminants holds significant potential for advancing air pollutant monitoring technologies, providing a pathway for more effective and portable detection. This advancement can contribute to the necessary alert actions for minimizing contaminants impacting public and environmental health, thereby safeguarding human and ecosystem health.

\end{abstract}

\begin{graphicalabstract}
\includegraphics[width=\linewidth]{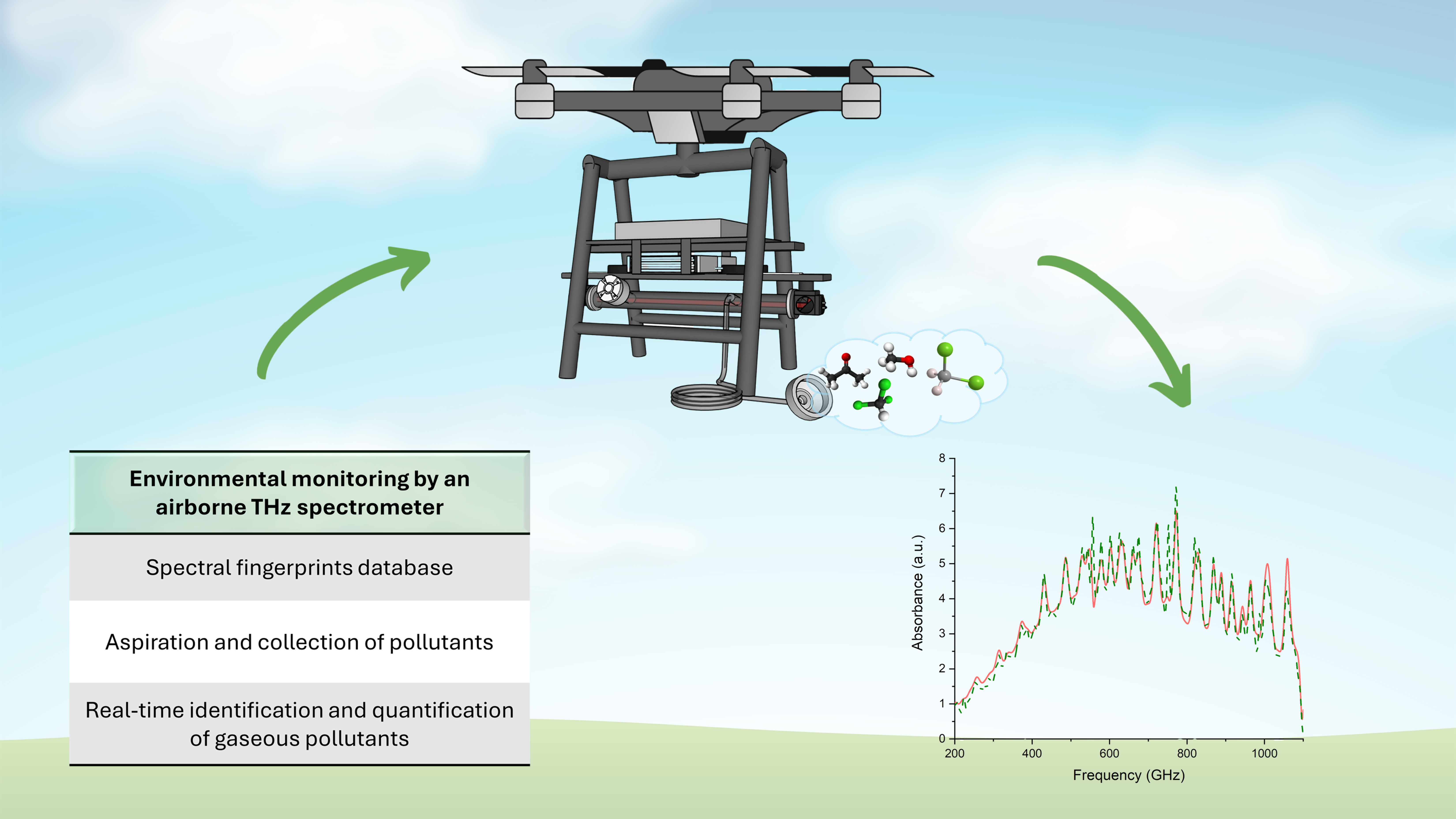}
\end{graphicalabstract}

\begin{highlights}
\item A terahertz-on-drone detector based on CW laser technology is developed for detect environmental pollutants.
\item The proof-of-concept system is not affected by flight vibrations and can detect the spectral fingerprints of pure gaseous pollutants in good agreement with laboratory data and previous scientific studies.
\item The set-up is engineered with an aspiration system that can perform real-time detection in flight above a source of pollutants and detect a multi-component mixture of environmental atmospheric contaminants.
\end{highlights}

%% Keywords

\end{frontmatter}



\section{Introduction}
\label{sec1}
%Addressing environmental pollution through responsible stewardship is increasingly recognized as vital to protecting both ecological systems and human well-being, this study present a potential groundbreaking compact and portable continuous-wave (CW) laser-based terahertz (THz) airborne sensor. 

Addressing environmental pollution through responsible stewardship is increasingly recognized as vital to protecting both ecological systems and human well-being. In this context, we present a potentially groundbreaking compact, and portable continuous-wave (CW) laser-based terahertz (THz) airborne sensor.

Terahertz spectroscopy represents an innovative non-invasive analytical tool in several fields including security screening \cite{Ikari2023, Kumar2023, Singh2024, Witko2012}, medical and biological investigations \cite{Zhan2023, Chernomyrdin2023, Yang2023, Gezimati2023, Nourinovin2023, Wu2023}, and cultural heritage diagnostic \cite{Kleist2019, Kleist2019(2), Moffa2024, Moffa2025, Moffa2025(2), Fukunaga2023, Fukunaga2024, Moffa2024(2), Moffa2024(3)}. 
The THz frequency range (0.1–10 THz) probes the low-frequency vibrational and rotational transitions of numerous molecules. The low photon energies within this range (4.2 meV at 1 THz) minimize the risk of molecular ionization and preclude combustion of flammable materials, rendering THz spectroscopy safe for biological samples \cite{Naftaly2015} and enabling remote sensing in populated environments.
\noindent
Furthermore, THz spectroscopy holds significant potential for atmospheric monitoring and environmental pollutant detection. 
\noindent
Many atmospheric molecules exhibit distinct rotational absorption lines in the THz spectral region \cite{Bigourd2006, Hindle2008, Galstyan2021, DArco2022, Tyree2022, Wang2022, Passarelli2022, Powers2023}, facilitating selective identification. Compared to infrared spectroscopy, THz measurements are less susceptible to interference from atmospheric particulates, fog, smoke, or dust  \cite{Bigourd2006, Su2012, Demers2017, Rice2021}. This reduced scattering arises from the THz wavelengths, which are significantly larger than typical atmospheric particle sizes. Notwithstanding these benefits, the practical implementation of real-time, field-deployable THz spectrometers for gas analysis has been largely restricted to humidity sensing, with limited evidence of detection in realistic field conditions or integration of gas cells and aspiration mechanisms \cite{Demers2017, Demers2018}. 

The proof-of-concept device presented in this work marks a significant leap forward in in situ atmospheric pollutant analysis, offering an unprecedented tool for safeguarding our planet by environmental monitoring.
This innovative prototype is a self-contained system, integrating a custom spectral library and a dedicated quantification algorithm. This synergistic design demonstrates the inherent feasibility of remote environmental monitoring  establishing a new benchmark for actionable environmental intelligence. We have validated the sensor's capacity for efficient collection, identification, and quantification of multi-component gas mixtures. The compelling results underscore the immense potential of drone-deployable THz detectors for providing reliable, real-time identification and quantification of gaseous pollutants.
This synergistic between laser and drone technologies lays the foundation for a robust environmental monitoring tool, directly contributing to pollution control and eventually ecological preservation. Its innovative approach to pollutant analysis is particularly impactful in geographically challenging locations or where physical obstructions impede real-time measurements of anthropogenic emissions. For example, deploying this system in remote industrial zones, over vast agricultural lands, or even within urban canyons can pinpoint pollution sources with unparalleled precision.
The prototype presented in this work is not only a scientific demonstration, it is also a springboard for future advancements. It can be further engineered to optimize its performance for specific real-life applications. For instance, reducing the system’s dimensions and weight would enhance its versatility for drone deployment, while the implementation of a multi-pass absorption cell (as discussed throughout the paper) would dramatically increase the system's sensitivity. This work represents a significant novel approach by presenting practical real-field data acquisition, analysis, and interpretation, which complements and expands upon previous methodologies (including the pioneering work \cite{Moffa2025_arxiv}).This advancement is a key step for developing future technologies that will empower us to better monitor, mitigate and possibly understand the environmental challenges of our time.

%, ultimately contributing to a healthier planet and improved welfare for all.
%This study presents a proof-of-concept compact and portable continuous-wave (CW) laser-based terahertz (THz) sensor for in situ analysis of atmospheric pollutants. The self-contained prototype integrates a custom spectral library and a dedicated quantification algorithm, demonstrating the feasibility of remote environmental monitoring. We experimentally validate the sensor's capacity for efficient collection, identification, and quantification of multi-component gas mixtures. The obtained results demonstrate the potential of drone-deployable THz detectors for reliable, real-time identification and quantification of gaseous pollutants. This technology offers a basis for a robust tool for environmental monitoring, contributing to pollution control and ecological preservation. The proposed system explores an innovative approach to pollutant analysis in geographically challenging locations or where physical obstructions impede real-time measurements of anthropogenic emissions.
%Furthermore, the system's dimensions and weight can be minimized through design optimizations, such as employing a multi-pass absorption cell to reduce the optical path length,advancing the development of future technology. 

\section{Materials and methods}
\label{sec2}

\subsection{Terahertz-on-drone system}
\label{ToD}
\begin{figure} [h!]
    \centering
    \includegraphics[width=\linewidth]{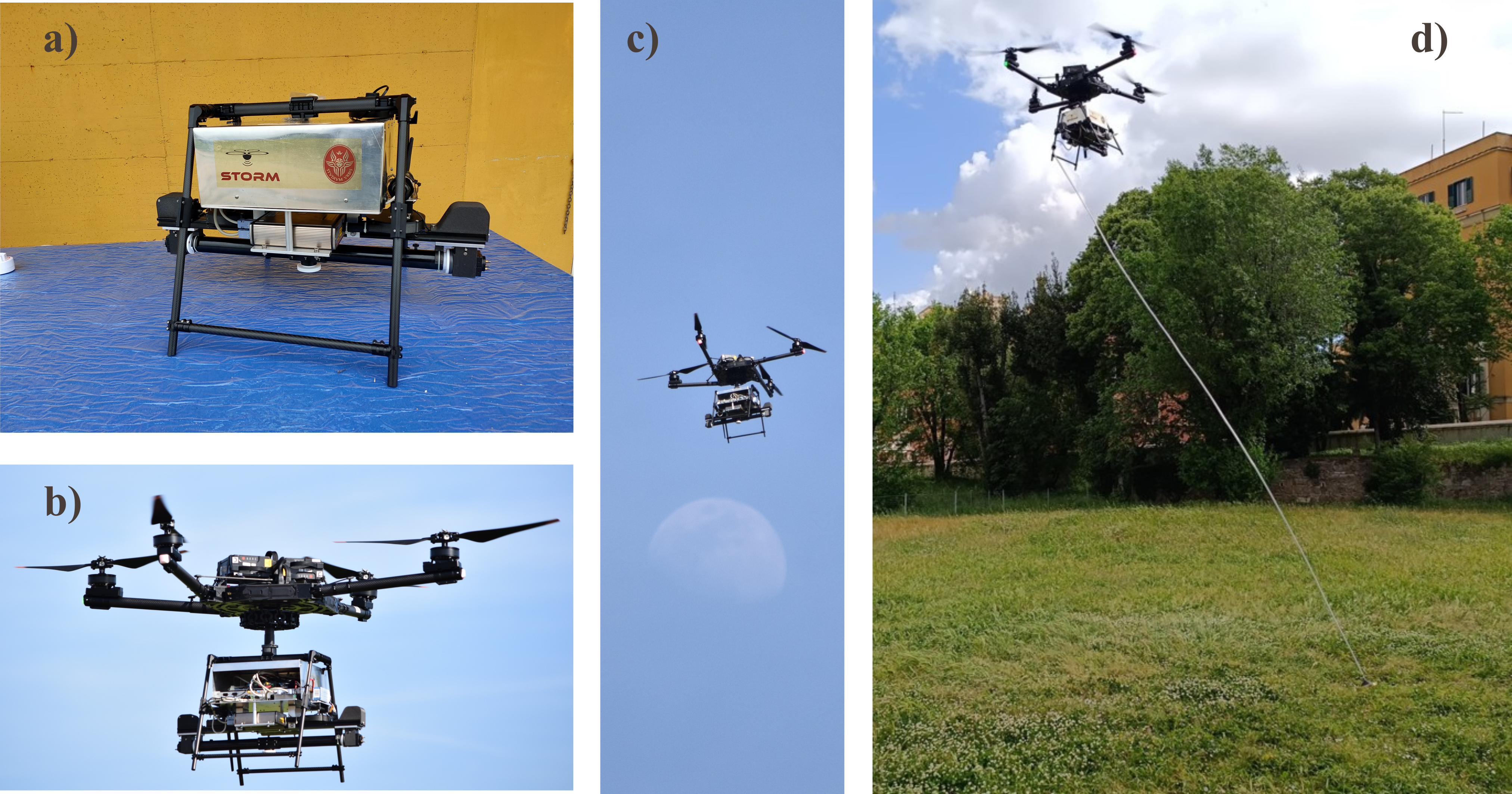}
    \caption{a) The assembled UAS-THz-CW system in ground station; b) and c) in flight during in-field trials for stability test; d) UAS-THz-CW system in flight during remote sampling with the 5-meter aspiration tube for the collection of the targeted analyte.}
    \label{UAS-THz-CW}
\end{figure}

This study introduces a proof-of-concept airborne THz-CW spectrometer for atmospheric pollutant detection (Figure \ref{UAS-THz-CW}). This system is engineered for robust performance, achieved by minimizing weight and power consumption, which in turn enhances stability and reduces vibrations during aerial deployment.
%This work presents a proof-of-concept drone-integrated terahertz continuous-wave (THz-CW) spectroscopic system (Figure \ref{UAS-THz-CW}) engineered to achieve robust performances while minimizing weight and power consumption to enhance stability and reduce vibrations.
The platform utilizes a Freefly Alta X unmanned aerial system (UAS), selected for its 31-minute flight autonomy with a payload of 5.5 kg and two 16 Ah 44.4V batteries. The maximum load in this condition is 15.06 kg in safety and the system is supported by an efficient anti-vibration mechanism capable of eliminating annoying mechanical disturbances towards the onboard instrumentation anchored. Centimeter-precision RTK GPS and vibration-dampened rotors, reducing typical rotor-induced vibrations by a factor of five, further enhance system stability. The UAS is remotely controlled via TCP/IP protocol using a high-gain omnidirectional antenna operating in the 2.4 GHz or 5 GHz WiFi bands, enabling real-time data transfer and instrument control.

\begin{figure}
    \centering
    \includegraphics[width=\linewidth]{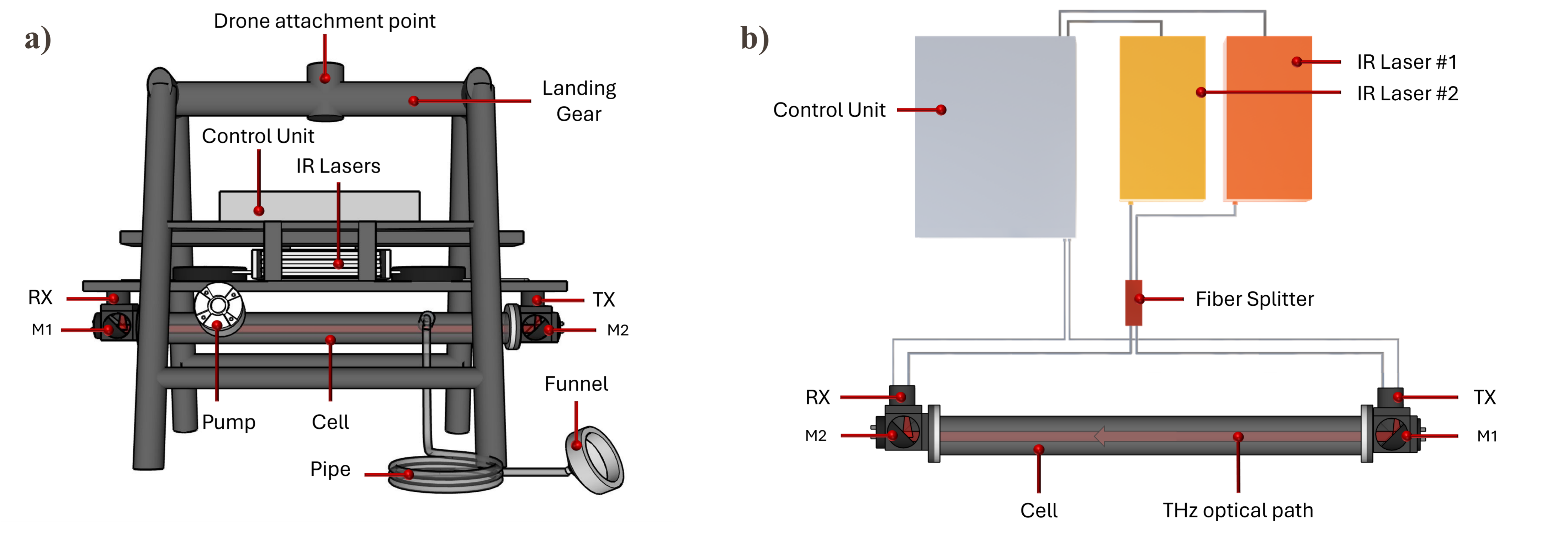}
    \caption{Schematic representations of: a) the assembled THz-on-drone system; b) CW spectrometer he radiation's optical path inside the sampling cell is reported in red.}
    \label{scheme}
\end{figure}

THz radiation is generated from the heterodyne downconversion of two infrared distributed feedback lasers (DFB, with a weight of 1 kg for each module) achieving a high-frequency resolution (in the order of 10 MHz).
The laser light irradiates the photomixer ($InGaAs/InP$) at two adjacent frequencies. Applying a bias voltage to the metal electrodes then generates a photocurrent that oscillates at the beat frequency. An antenna structure surrounding the photomixer translates the oscillating photocurrent into the terahertz wave.  
The fiber-coupled system is designed for compactness and low weight to optimize flight time and thrust-to-weight ratio of the THz-on-drone detector.
The CW lasers' combination generates single THz frequencies and when dynamically tuned covers the range between 0.1 and 1.1 THz.
The system's motherboard is positioned on the central line of the landing gear connected to Alta X which is made of carbon fiber, with a bell-shaped structure. To anchor and give greater structural stability to the landing gear, two horizontal carbon tubes are installed on the sides, fixed using 3D-printed corner supports, on which an aluminum plate made by laser cutting is placed.
The fiber-based transmitter (TX) and receiver (RX) of the THz-CW system (photoconductive antennas, PCA) are placed in two 3D-printed ogives at the ends of a 70 cm aluminum track. Off-axis parabolic mirrors (50.8 mm focal length) within the ogives collimate the THz radiation (as schematically represented in Figure \ref{scheme}). The ogives protect the optical fibers from aerodynamic forces during flight.

A 54 cm long, 32 mm inner diameter carbon fiber cylindrical sampling cell, fitted with two 3 cm diameter, 5 mm thick Teflon windows (transparent to THz radiation) perpendicular to the beam propagation, is positioned between the PCAs. The system incorporates a 12 L/min, 12 V pump and a 5-meter aspiration pipe (6.5 mm inner diameter) with a wide-opening funnel for gaseous sample collection. An automated control system regulates pump operation and valve closure for precise sampling. All system components are battery-powered.

\subsection{Measurements protocol}
\label{mixSec}

%A comprehensive spectral database of gaseous pollutants in the terahertz (THz) frequency range was meticulously constructed under controlled laboratory conditions. Pure compounds were introduced into an evacuated sampling cell (P=100 mbar), with the empty cell spectrum serving as the reference. Precise monitoring of each compound's partial pressure during measurements enabled accurate determination of optical parameters. A complementary database, generated with the same compounds in ambient air, exhibited excellent agreement with the vacuum dataset.
%To evaluate potential pressure-induced effects on the spectral line shapes, a systematic comparison was performed between measurements acquired under vacuum conditions and those obtained at ambient atmospheric pressure. The results revealed no appreciable broadening or distortion of the spectral features, indicating that intermolecular collisions and pressure variations exert a negligible influence within our THz spectroscopic setup. Importantly, this ensures that the quantification of analytes remains unaffected by the measurement environment.
A comprehensive spectral database of gaseous pollutants in the terahertz (THz) frequency range was meticulously constructed under controlled laboratory conditions. For each pure compound, it was introduced into an evacuated sampling cell (P = 100 mbar), with the empty cell spectrum serving as a precise reference. During these measurements, each compound's partial pressure was monitored, enabling the accurate determination of optical parameters. To assess any potential pressure-induced effects on spectral line shapes, measurements acquired under vacuum conditions were compared with those obtained at ambient atmospheric pressure, thereby generating a new database for pure-in-air fingerprints. A slight broadening of the spectral features for some compounds was observed compared to the vacuum dataset. Consequently, data analysis initially utilized only the pure-in-vacuum fingerprints and then the pure-in-air, with the comparison revealing a negligible effect on data interpretation. 
The validation of the prototype's functionality was divided into two main phases: Phase 1: Ground-based verification and Phase 2: In-flight verification. These phases described in detail in Section \ref{res} and, are briefly summarized here for clarity.

\subsection{Phase 1: Ground-Based Verification}
\label{Phase1}

In Phase 1, building on foundational data, the integrated drone-mounted THz system was initially evaluated at a ground station. This involved acquiring spectra with the drone held stationary at a precise location. In this initial phase, reagent-grade acetone ($CH_3COCH_3$, $\geq$ 99.5 \%, CAS No.: 67-64-1), methanol ($CH_3OH$, $\geq$ 99.8\%, CAS No.: 67-56-1), acetonitrile ($CH_3CN$, $\geq$ 99.93 \%, CAS No.: 75-05-8), ammonia (in aqueous solution $NH_4OH$, $\sim$ 30 \%, CAS No.: 1336-21-6), and dichloromethane ($CH_2Cl_2$, $\geq$ 99.8\%, CAS No.: 75-09-2), all procured from Sigma-Aldrich, were introduced into the sampling cell. Absorption spectra were obtained following the methodology described in the following (Equations \ref{abs_ToD} and \ref{alpha_ToD}) and compared with previously acquired laboratory data and literature values. Additionally, gas mixtures of these compounds was aspirated into the gas chamber and analyzed to ensure the aspiration system's effectiveness prior to its use during Phase 2. These steps and results served to validate, by comparison, the drone-integrated system's ability to detect these analytes during in-flight operation in Phase 2.
Prior to characterizing any gaseous compound ($I_{sample}$), a reference spectrum ($I_{reference}$) of ambient air was always acquired; this reference was critical for subsequent data analysis, particularly considering water vapor content. All measurements were performed with 100 MHz spectral resolution and a 10 ms integration time per data point. This optimized acquisition strategy enabled rapid spectral collection without compromising resolution, a crucial factor for efficient field deployment.
Raw data were pre-processed using a custom algorithm implemented in MATLAB (ver. 2019a, MathWorks Inc., USA). The envelope of the raw photocurrent signal was extracted using the Hilbert transform method \cite{Kong2018, Vogt2019, DArco2022, Moffa2024, Moffa2024(2)}. 
The amplitude was then transformed into the time domain via fast Fourier transform (FFT), filtered to remove spurious oscillations, and subsequently converted back to the frequency domain. This procedure was applied to both the reference and sample spectra. The experimental absorbance, $A(\nu)$, was calculated according to the Beer-Lambert Law as follows:

\begin{equation}
\centering
A(\nu)=-log\frac{I_{sample}}{I_{reference}}=-log(T(\nu))   
\label{abs_ToD}
\end{equation}

where $A(\nu)$ is the absorbance as a function of frequency, $I_{sample}$ and $I_{reference}$ are the energy spectral densities of reference and sample and $T(\nu)$ is the experimental transmittance.

Subsequently, molar absorption coefficient $\alpha(\nu)$ was derived from the absorbance according to:

\begin{equation}
\centering
\alpha(\nu) = \frac{A(\nu)}{c \cdot l}
\label{alpha_ToD}
\end{equation}

where $c$ is the molar concentration of the analyte (in mol·L$^{-1}$=M), and $l$ is the optical path length of the sampling cell (in cm). The resulting $\alpha(\nu)$ is thus expressed in units of M$^{-1}$·cm$^{-1}$, enabling direct comparison of absorption intensities independent of sample concentration or path length.

%Building on foundational data, we initially evaluated the integrated drone-mounted THz system at a ground station (PHASE1). This involved acquiring spectra with the drone held stationary at the precise location. Following successful validation of this ground-based data, we proceeded to conduct in-flight measurements, as detailed in Section 3.
%For flight tests (PHASE2), we employed aspiration-based sampling to analyze individual gaseous compounds and their mixtures. During sampling, an aspiration pump was activated and valves were opened to draw in the analyte into the gas chamber on the drone. Once sampling was complete, the pump was deactivated and the valves were closed, preparing the system for spectral analysis of the collected analytes.
%Prior to characterizing the gaseous compound, a reference spectrum of ambient air was always acquired; this reference was critical for subsequent data analysis taking into consideration water vapors. All measurements were performed with 100 MHz spectral resolution and a 10 ms integration time per data point. This optimized acquisition strategy enabled rapid spectral collection without compromising resolution, a crucial factor for efficient field deployment. 
%The validation of the prototype's functionality can be divided into two main phases: PHASE1: "Ground-based" verification and PHASE2: "In-flight" verification. These phases are described in detail in the following paragraph \ref{res} and briefly summarized below for clarity.

\subsection{Phase 2: In-Flight Verification}
\label{Phase2}

Following successful validation of the ground-based data, in-flight measurements were conducted (Phase 2), as detailed in Section \ref{res}. For flight tests, aspiration-based sampling was employed to analyze individual gaseous compounds and their mixtures. During sampling, an aspiration pump was activated, and valves were opened to draw the analyte into the gas chamber on the drone. Once sampling was complete, the pump was deactivated and the valves were closed, preparing the system for spectral analysis of the collected analytes.
In Phase 2, two multi-component gas mixtures were investigated. Firstly, a liquid sample containing acetone, methanol, and dichloromethane was prepared and placed in a sample holder at the ground level. 
Lastly, a more complex liquid sample containing ammonia, acetonitrile, acetone, methanol, and dichloromethane was prepared and placed in a sample holder at the ground level. The emitted gaseous analytes were then aspirated into the sampling cell of the drone-mounted prototype during flight. Spectral analysis of the acquired signal was performed using a multiple absorber approach, assuming non-interacting components, and compared against an in-house developed THz spectral database of pure gaseous compounds according to:

\begin{equation}
     A_{Retrieved}(\nu)=\sum_{i=1}^{n} \alpha_{i}(\nu)\cdot x_i
     \label{mix}
\end{equation}

where $\alpha_{i}$ represents the molecular absorption coefficient of the pure substances and $x_i$ represents the weighted coefficient for each compound expressed in $M\cdot cm$.
The mixture spectrum is reconstructed from a weighted linear combination of pure spectra, and the coefficients allow to retrieve the quantity of each compound present in the mixture.

\begin{figure}[h!]
    \centering
    \includegraphics[width=\linewidth, height=0.7\textheight, keepaspectratio]{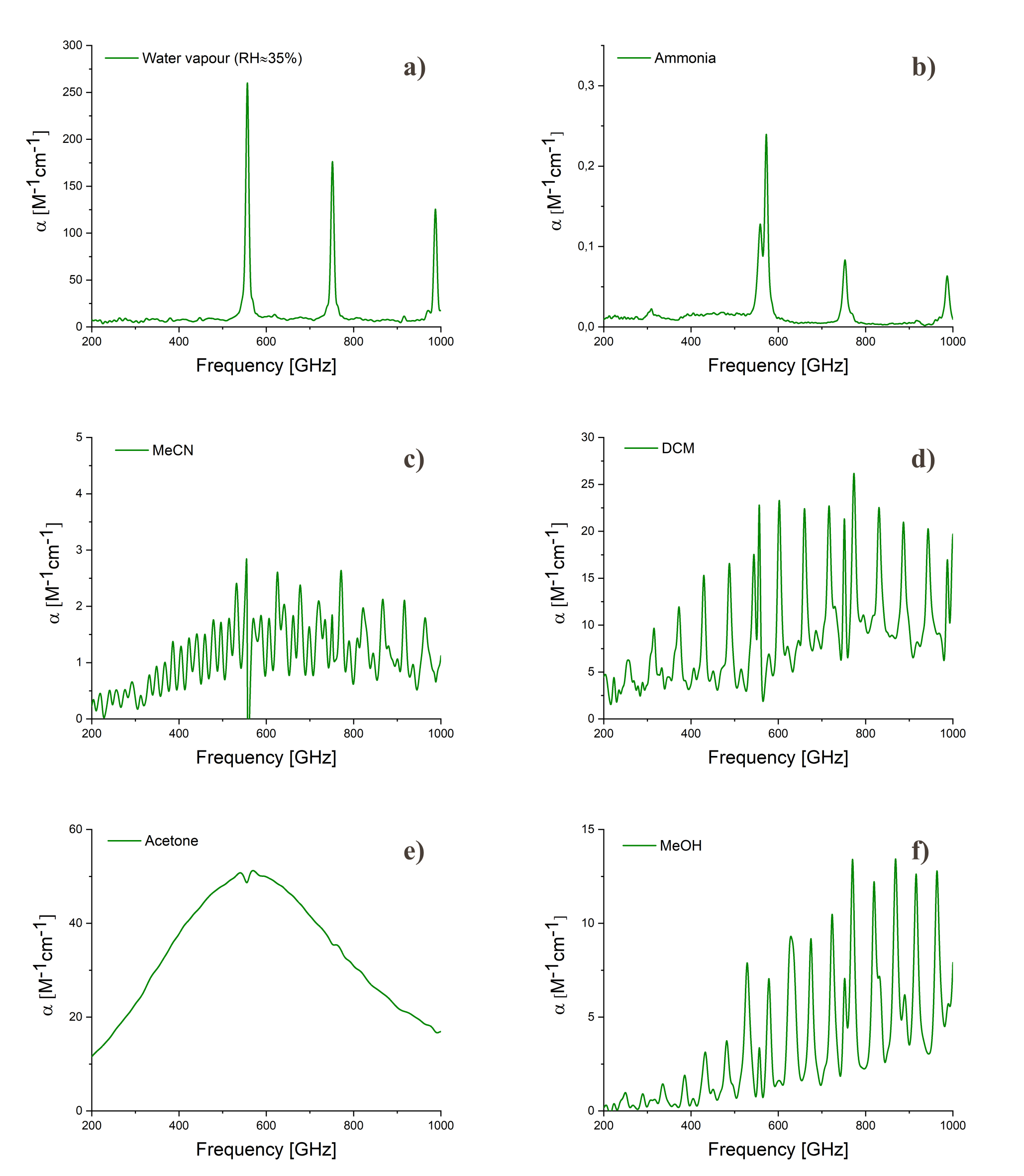}
    \caption{a) Absorption spectrum of water vapor (acquired in ambient air with a relative humidity (RH) of $\approx$ 35\%). The peaks at 557.2, 752.3 and 987.9 GHz, linked to water vapour absorptions \cite{Slocum2013}.
    b) Absorption spectrum of ammonia which presents a specific absorption fingerprints centered at 0.572 THz.
    c) Absorption spectrum of acetonitrile (MeCN) with specific fingerprints spaced approximately every 18.43 GHz.
    d) Absorption spectrum of dichloromethane (DCM) with specific fingerprints spaced approximately every 57.46 GHz. 
    e) Absorption spectrum of acetone injected into the cell wich presents a broad band centered at 0.565 THz. f) Absorption spectrum of methanol (MeOH) with specific fingerprints spaced approximately every 49.36 GHz.
    All the measurements are carried out in ambient air in the range 0.2-1 THz. The water vapour absorption peaks are present at 557.2, 752.3 and 987.9 GHz \cite{Slocum2013} in every spectra.}
    \label{CW_pure}
\end{figure}

\section{Results and discussion}
\label{res}

To evaluate the feasibility of our prototype for atmospheric pollutant detection, we conducted a series of rigorous experiments. To the best of our knowledge, this study marks the first demonstration of a drone-mounted THz sensor incorporating a gas chamber and an aspiration system for remote, point-of-care air sampling.

\begin{figure}[h!]
    \centering
    \includegraphics[width=\linewidth, height=0.7\textheight, keepaspectratio]{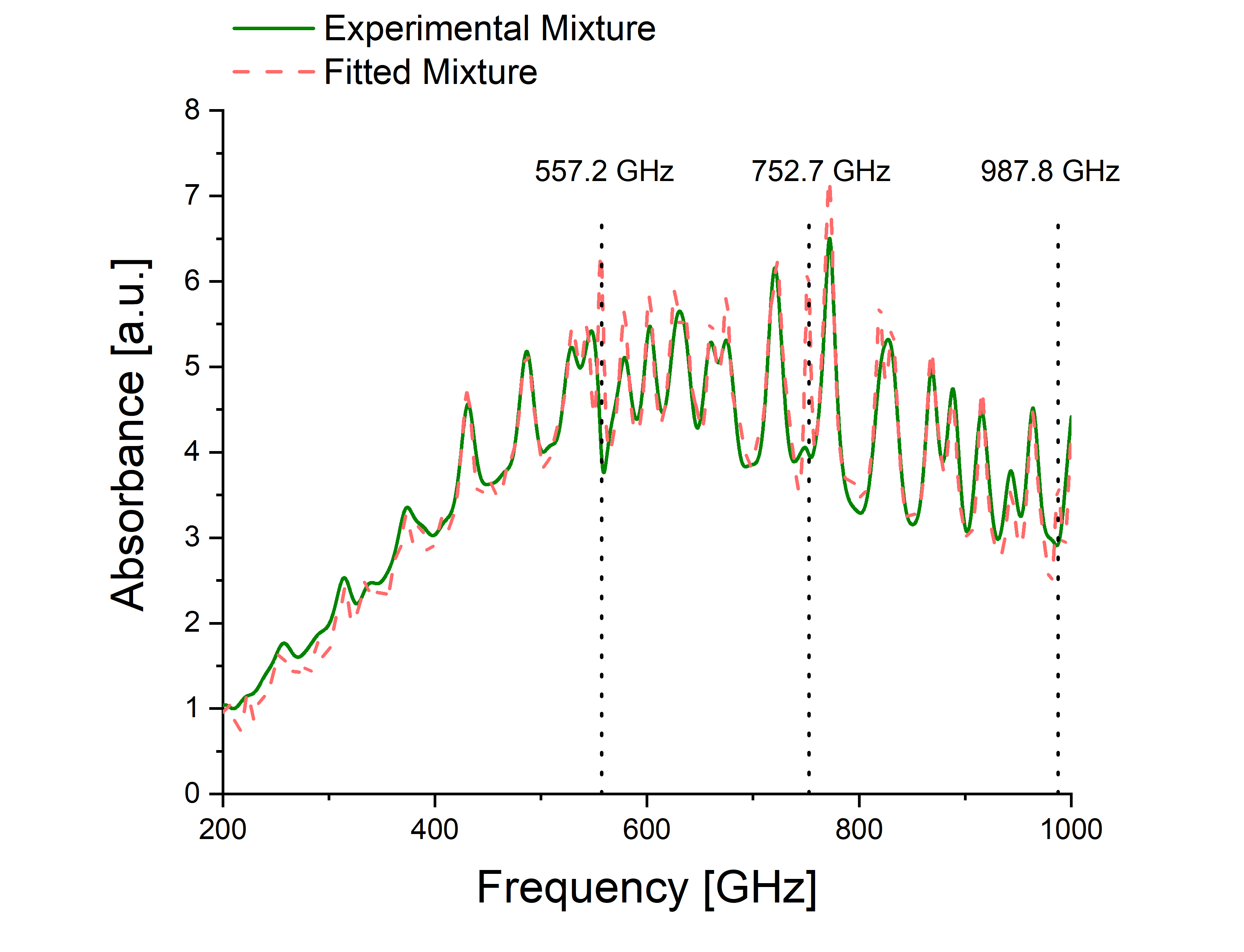}
    \caption{Comparison between the experimental mixture constituted by acetone, methanol, and dichloromethane aspirated inside the sampling cell and remotely investigated by the THz-on-drone system at a flight altitude from the soil level of 5 m and the spectrum obtained from the retrieved coefficients in the range 0.2-1 THz. Dotted vertical lines indicate water vapour absorptions frequencies (557.2, 752.3 and 987.9 GHz) \cite{Slocum2013}.}
    \label{Mix}
\end{figure}

Initial flight tests were performed to assess the system's stability and vibration resistance during a 15-minute flight at an altitude of 100 meters. Data acquired at 500.7 GHz, with a 30 ms integration time and 100 MHz resolution, yielded a signal stability of approximately 1\% standard deviation, consistent with ground-based measurements. This critical finding confirmed the system's robustness under dynamic flight conditions.

Subsequently, we performed spectral measurements at varying altitudes to analyze gaseous compounds from an in-house environmental pollutant database. To validate the system's accuracy, we introduced pure samples of acetone, methanol, acetonitrile, ammonia and dichloromethane into the sampling cell and compared the acquired spectra with existing laboratory data. 
%The acetone spectrum (Fig. \ref{CW_pure}a) obtained at 100 meters altitude exhibited the characteristic broad absorption band at 0.565 THz, consistent with literature. Similarly, the methanol (Fig. \ref{CW_pure}b) and dichloromethane (Fig. \ref{CW_pure}c) spectra revealed distinct rotational absorption lines, aligning precisely with both our database and published literature values. 
%The obtained spectra (Figure \ref{CW_pure}), revealed distinct rotational absorption lines, aligning precisely with both our database and published literature values.
The obtained spectra (Figure \ref{CW_pure}) exhibited characteristic absorption features-ranging from distinct rotational lines to single broad bands—in excellent agreement with both our reference database and published literature values.
This result demonstrates the system's ability to accurately identify these volatile organic compounds (VOCs) during flight. Notably, the dichloromethane fingerprints represent the first THz spectroscopic study of this compound that matches theoretical expectations, as detailed in our work \cite{Moffa2025_arxiv}.

\begin{figure}[h!]
    \centering
    \includegraphics[width=\linewidth, height=0.7\textheight, keepaspectratio]{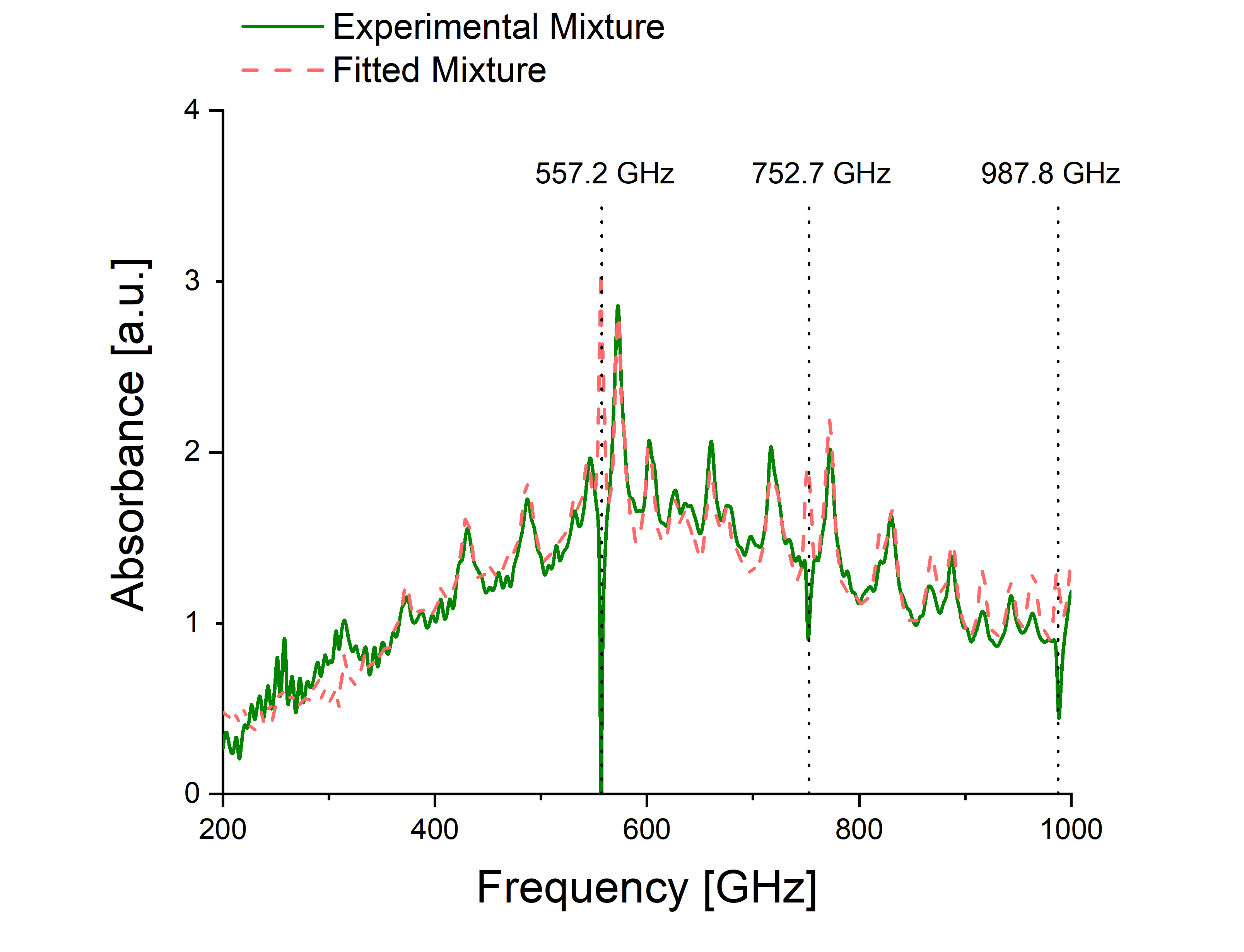}
    \caption{Comparison between the experimental mixture constituted by ammonia, acetonitrile, acetone, methanol, and dichloromethane aspirated inside the sampling cell and the spectrum obtained from the retrieved coefficients in the range 0.2-1 THz. Dotted vertical lines indicate water vapour absorptions frequencies (557.2, 752.3 and 987.9 GHz) \cite{Slocum2013}.}
    \label{Mix5}
\end{figure}

We evaluated the system's unique remote sampling capabilities using an integrated 5-meter tube and funnel, which enabled targeted sampling at potential pollutant sources. To further demonstrate the system's quantitative accuracy, we prepared a liquid mixture containing acetone, methanol, and dichloromethane. This mixture was then placed in a sample holder on a grass field. A reference spectrum of ambient air was acquired at 5 meters altitude, followed by sample collection and spectral analysis using the multiple absorber approach (Section \ref{Phase2}). The experimental spectrum was then fitted using a linear combination of pure compound spectra (Figure \ref{Mix}), which accurately identified the three components and allowed quantification of their presence in the mixture: $3.64 \cdot 10^{-6}$ mol (acetone), $5.95 \cdot 10^{-4}$ mol (dichloromethane), and $8.65 \cdot 10^{-4}$ mol (methanol). 
A second more chemically complex liquid mixture (Figure \ref{Mix5}), consisting of ammonia, acetonitrile, acetone, methanol, and dichloromethane, was prepared, with the sample holder once again placed in direct contact with the surface of the ground. Despite the increased spectral overlap and molecular diversity, the individual concentrations of each compound aspirated into the sampling cell were successfully determined. This was accomplished using the proposed spectral fitting approach, which enabled effective deconvolution of the absorption features and a robust retrieval of the analytes' concentrations within the mixture, yielding values of \(5.99 \cdot 10^{-2}\) mol for ammonia, \(1.78 \cdot 10^{-4}\) mol for methanol, \(3.46 \cdot 10^{-4}\) mol for dichloromethane, \(2.89 \cdot 10^{-4}\) mol for acetone, and \(2.97 \cdot 10^{-4}\) mol for acetonitrile.

In the presented results, the humidity fluctuations between the reference and the sample measurements appear as positive or negative peaks (depending on if the contribution of water vapor in the sample is higher or lower than the reference), these effects do not compromise the identification of the analytes. Additionally, the consistent presence of water vapor absorption features throughout the measurements serves as a useful internal reference to confirm the proper functioning and stability of the spectrometer.
We chose to retain the water vapor features in the spectra because they do not significantly interfere with the absorption bands of the targeted compounds. Even in cases of partial overlap, the analytes can still be reliably identified due to the distinct absorption patterns and spectral structures that are characteristic of each substance. 
While it is conceivable for an analyte to possess a single absorption peak that coincidentally overlaps in both frequency and linewidth with a specific water vapor line, such occurrences are highly improbable. Should this occur, accurate quantification of that particular analyte would necessitate precise knowledge of the ambient humidity, complicating direct detection. This highlights the indispensable role of comprehensive spectral databases for pure analytes, as exemplified by our work. Ultimately, the inherent differences in their THz absorption characteristics confer high selectivity upon this spectroscopic technique for environmental sensing.

As illustrated by these findings, this data analysis provides a valuable measure of the relative amounts of substances captured within the gas chamber in real-world field applications. Although dynamic environmental factors such as wind and source proximity can influence the specific collected concentrations, the key achievement is the system's demonstrated ability to reliably identify and quantify the substances present in the sampled air. The high-resolution drone-mounted THz sensor enabled selective measurements, resolving individual components within the gas mixture even under real ambient conditions. Following the validation of the system and its ability to detect an increased number of compounds, the development of a comprehensive spectral database covering all relevant pollutants will advance the applicability of this proof-of-concept.

%We evaluated the system's unique remote sampling capabilities using an integrated 5-meter tube and funnel, which enabled targeted sampling at potential pollutant sources. To further demonstrate the system's quantitative accuracy, we prepared a liquid mixture containing acetone, methanol, and dichloromethane (20\%, 40\%, and 40\% by volume, respectively, in a very small total volume). This mixture was then placed in a sample holder on a grass field.
%A reference spectrum of ambient air was acquired at 5 meters altitude, followed by sample collection and spectral analysis using the multi-absorber approach (Section \ref{ToD}). The experimental spectrum was then fitted using a linear combination of pure compound spectra, which accurately identified the three components and allowed for the quantification of their presence in the mixture: $3.64 \cdot 10^{-6}$ mol (acetone), $5.95 \cdot 10^{-4}$ mol (dichloromethane) and $8.65 \cdot 10^{-4}$ mol (methanol). As illustrated by these findings, this data analysis provides a valuable measure of the relative amounts of substances captured within the gas chamber in real-world field applications. While dynamic environmental factors like wind and source proximity can influence the specific concentrations collected, the key achievement is the system's demonstrated ability to reliably identify and quantify the substances present in the sampled air. The high-resolution drone-mounted THz sensor enabled selective measurements, resolving individual components within the gas mixture even under real ambient conditions.

Sensor sensitivity was assessed by correlating the intensity of the absorption peak with the system's noise, defined as the standard deviation ($\sigma$) of the signal with a blocked THz beam \cite{Neumaier2015, Hepp2016}. 
%At a 30 ms integration time and a 1-meter optical path, detection sensitivities are estimated to be approximately in the range $10^{-6}-10^{-4}$ $M$, corresponding respectively to the most and least sensitive analytes examined in this study.
For an integration time of 30 ms and an optical path length of 1 meter, the estimated detection sensitivities range from $ \sim 10^{-6}$ M to $\sim 10^{-4}$ M, corresponding to the most and least responsive analytes investigated in this study, respectively.

%These limits can be further reduced by engineering system performance; for instance, by modifying the gas chamber of our setup to implement a 10-meter multipass gas cell, simulations indicated that the detection limits could be significantly improved to $ \approx 10^{-5} - 10^{-3}$ $M$. Previous studies have demonstrated that by employing also a gas concentrator the sensitivities can go further down to $10^{-11}-10^{-8}$ $M$ for common VOCs \cite{Neese2012, Fosnight2013, Neumaier2015, Hepp2016}.  

These limits can be further reduced by optimizing system performance. For example, simulations indicate that modifying the current gas chamber to incorporate a 10-meter multipass configuration could significantly enhance sensitivity, lowering detection limits to  $ \sim 10^{-7}$–$10^{-5}$ M. Furthermore, previous studies have shown that the addition of a gas concentrator can increase detection thresholds even further, achieving sensitivities in the range of $10^{-11}$–$10^{-8}$ M for common volatile organic compounds (VOCs) \cite{Neese2012, Fosnight2013, Neumaier2015, Hepp2016}.

Intrinsic spectrometer parameters, such as integration time and spectral range, significantly influence sensitivity. Although increasing integration time enhances the signal-to-noise ratio (SNR), this must be carefully balanced against drone autonomy and the requirements for spatial gas concentration mapping. Expanding the usable spectral range can improve the sensitivity of specific analytes. For instance, in the case of gaseous methanol, the limit of detection is expected to be lower within the 1-2 THz range due to its increased absorbance in this region, thereby allowing the detection of smaller concentrations of methanol and improving the system's sensitivity within this specific frequency range.

This proof-of-concept study demonstrates the feasibility of the airborne THz spectrometer for environmental monitoring. It successfully achieves a crucial balance between measurement speed, spectral range, and sampling cell availability. The innovative integration of an aspiration system with drone-based THz spectroscopy, as presented here, constitutes a novel and powerful tool for remote, point-of-measurement air quality monitoring. This system enables efficient detection and quantification of gaseous pollutants and holds significant potential for further optimization across a wider range of applications, paving the way for more effective environmental stewardship.

\section{Conclusions}

This study introduces a proof-of-concept drone-integrated terahertz (THz) system designed for high-resolution, remotely controlled in-field spectroscopic analysis. Our portable and lightweight prototype successfully demonstrates its capability for detecting atmospheric gaseous pollutants.
Notably, this work presents the first-ever in-flight continuous-wave (CW) THz characterization of pure polluting compounds and their mixtures, representing a significant novel contribution to the field. The acquired spectra exhibit excellent agreement with laboratory-based measurements, thoroughly validating the system's inherent stability and resilience to drone-induced vibrations. Furthermore, the system incorporates a unique remote sampling capability, effectively employing a pump and a 5-meter aspiration tube for the targeted collection of analytes.

All measurements in this study were conducted under standard ambient atmospheric conditions (T = 298 K, P = 1 atm), consistent with the intended operational environment of the developed prototype. Minor environmental variations, such as those arising from seasonal shifts or geographical differences, are not expected to substantially affect the absorption spectra in a manner that would compromise system performance. In scenarios where the system is deployed in environments with significantly different ambient conditions that lead to not negligible alterations in line shapes, system database recalibration with newly acquired reference spectra would be necessary \cite{Cheville1995, Cheville1999, Matton2006, Bigourd2007, Cazzoli2008, Cazzoli2009, Svelto2010, Ren2010}.

We demonstrated the prototype functionality by successfully acquiring and analyzing a multi-component pollutant mixture emitted from a ground-based source during a 5-meter altitude flight. The system accurately identified and quantified the individual components within the mixture, showcasing its immense potential for real-time, on-site environmental analysis. This is particularly crucial in scenarios where ground-based access is difficult or hazardous.

%This research highlights the transformative potential of combining THz-CW spectroscopy with unmanned aerial systems (UAS) to detect and analyze multicomponent pollutants. Further expansion of the spectral fingerprint database for a broader range of contaminants will be pursued in future work to enhance detection accuracy and system applicability across diverse environmental conditions.

This research highlights the transformative potential of combining THz-CW spectroscopy with unmanned aerial systems (UAS) for environmental monitoring. The ability to perform high-resolution, remotely controlled measurements offers a powerful tool for real-time identification and quantification of pollutants across diverse applications, ranging from natural environment contaminant detection to industrial emission monitoring. 
Crucially, the inherent simplicity and miniaturization potential of THz-CW technology, combined with the accessibility of UAS platforms, suggests that this approach could be scaled for cost-effective, widespread deployment. This enables comprehensive environmental monitoring even in remote or challenging terrains. Upgrading the system database with newer fingerprints of pollutants and any other substance of interest will increase the potential of the proof of concept presented in this work.

%This research highlights the transformative potential of combining THz-CW spectroscopy with unmanned aerial systems (UAS) for environmental monitoring. 

%The ability to perform high-resolution, remotely controlled measurements offers a powerful tool for real-time identification and quantification of pollutants across diverse applications, ranging from natural environment contaminant detection to industrial emission monitoring. 
%Crucially, the inherent simplicity and miniaturization potential of THz-CW technology, combined with the accessibility of UAS platforms, suggests that this approach could be scaled for cost-effective, widespread deployment. This enables comprehensive environmental monitoring even in remote or challenging terrains. 
This work paves the way for advanced environmental surveillance strategies, enabling rapid and spatially resolved assessments of air quality with unprecedented flexibility and precision.

\section*{Acknowledgements}

 This research was conducted in the framework of "STORM - Sensori su sistemi mobili e remoti al Terahertz PNRM a2017.153 STORM" funded by Ministero della Difesa and "R-SET: Remote sensing for the environment by THz radiation" Large Research Projects of Sapienza, University of Rome.
This work was carried out thanks to "PRIN 2022: TREX a prototype of a portable and remotely controlled platform based on THz technology to measure the one health vision: environment, food, plant health, security, human and animal health" funded by the European Union - Next Generation EU (CUP B53D23013610006 - Project Code 2022B3MLXB PNRR M4.C2.1.1). This work has been supported by PNRR MUR project
CN00000022 “Agritech” Spoke 9. This work was also supported by Sapienza competitive grants: Grandi Attrezzature Scientifiche (2018) titled "SapienzaTerahertz: THz spectroscopic image system for basic and applied sciences".  This work was supported by Sapienza Large Projects Research Call 2023 titled "TforCH: R\&D on the potentiality of THz radiation for Cultural Heritage". The authors would like to thank "CSN5 Grants INFN Roma 1" for their contribution to this work.

\section*{Author contributions statement}
\noindent Conceptualization, C.M. and M.P.; methodology, C.M. and M.P.; software, C.M., C.M., M.C. and M.P.; validation, C.M., A.C., C.M., V.M.O., D.F., F.Jr.P.M., L.G., M.M., G.Z., M.R., L.M., and M.P.; formal analysis, C.M., and M.P.; investigation, C.M., and M.P.; resources, M.M., M.C., L.G., M.M., G.Z., M.R., L.M., and M.P.; data curation, C.M., and M.P.; writing---original draft preparation, C.M. and M.P.; writing---review and editing, all the authors; visualization, C.M.; supervision, A.C., L.G., M.M., G.Z., M.R., L.M., and M.P.; project administration, C.M., and M.P.; funding acquisition, L.G., M.R., L.M., and M.P. \\
All authors have read and agreed to the published version of the manuscript.

\section*{Additional information}
\noindent
\textbf{Competing interests}: The authors declare no competing interests. 

\noindent
\textbf{Data availability}: The datasets used and/or analysed during the current study are available from the corresponding author on reasonable request.

%\appendix
%\section{Example Appendix Section}
%\label{app1}

\bibliographystyle{elsarticle-num} 
\bibliography{elsarticle/elsarticle/Ref}

\begin{thebibliography}{10}
\expandafter\ifx\csname url\endcsname\relax
  \def\url#1{\texttt{#1}}\fi
\expandafter\ifx\csname urlprefix\endcsname\relax\def\urlprefix{URL }\fi
\expandafter\ifx\csname href\endcsname\relax
  \def\href#1#2{#2} \def\path#1{#1}\fi

\bibitem{Ikari2023}
T.~Ikari, Y.~Sasaki, C.~Otani, 275--305 ghz fm-cw radar 3d imaging for walk-through security body scanner, in: Photonics, Vol.~10, MDPI, 2023, p. 343.

\bibitem{Kumar2023}
P.~N. Kumar, M.~Nagaraju, K.~Arjun, A.~Razdan, A.~Chaudhary, Comparative study of real-time terahertz imaging of concealed metallic objects, drug, wood and tnt explosive in transmission/reflection modes using uncooled microbolometer and ultrafast pulsed terahertz imaging systems, Indian Journal of Physics (2023) 1--10.

\bibitem{Singh2024}
K.~Singh, U.~Aalam, A.~Mishra, N.~Dixit, A.~Bandyopadhyay, A.~Sengupta, Spectroscopic and imaging considerations of thz-tds and ulf-raman techniques towards practical security applications, Optics Express 32~(2) (2024) 1314--1324.

\bibitem{Witko2012}
E.~M. Witko, T.~M. Korter, Terahertz spectroscopy of the explosive taggant 2, 3-dimethyl-2, 3-dinitrobutane, The Journal of Physical Chemistry A 116~(25) (2012) 6879--6884.

\bibitem{Zhan2023}
X.~Zhan, Y.~Liu, Z.~Chen, J.~Luo, S.~Yang, X.~Yang, Revolutionary approaches for cancer diagnosis by terahertz-based spectroscopy and imaging, Talanta (2023) 124483.

\bibitem{Chernomyrdin2023}
N.~Chernomyrdin, D.~Il’enkova, V.~Zhelnov, A.~Alekseeva, A.~Gavdush, G.~Musina, P.~Nikitin, A.~Kucheryavenko, I.~Dolganova, I.~Spektor, et~al., Quantitative polarization-sensitive super-resolution solid immersion microscopy reveals biological tissues’ birefringence in the terahertz range, Scientific Reports 13~(1) (2023) 16596.

\bibitem{Yang2023}
S.~Yang, L.~Ding, S.~Wang, C.~Du, L.~Feng, H.~Qiu, C.~Zhang, J.~Wu, K.~Fan, B.~Jin, et~al., Studying oral tissue via real-time high-resolution terahertz spectroscopic imaging, Physical Review Applied 19~(3) (2023) 034033.

\bibitem{Gezimati2023}
M.~Gezimati, G.~Singh, Advances in terahertz technology for cancer detection applications, Optical and Quantum Electronics 55~(2) (2023) 151.

\bibitem{Nourinovin2023}
S.~Nourinovin, M.~M. Rahman, S.~J. Park, H.~Hamid, M.~P. Philpott, A.~Alomainy, Terahertz dielectric characterisation of three-dimensional organotypic treated basal cell carcinoma and corresponding double debye model, IEEE Transactions on Terahertz Science and Technology (2023).

\bibitem{Wu2023}
X.~Wu, R.~Tao, T.~Zhang, X.~Liu, J.~Wang, Z.~Zhang, X.~Zhao, P.~Yang, Biomedical applications of terahertz spectra in clinical and molecular pathology of human glioma, Spectrochimica Acta Part A: Molecular and Biomolecular Spectroscopy 285 (2023) 121933.

\bibitem{Kleist2019}
E.~M. Kleist, C.~L. Koch~Dandolo, J.-P. Guillet, P.~Mounaix, T.~M. Korter, Terahertz spectroscopy and quantum mechanical simulations of crystalline copper-containing historical pigments, The Journal of Physical Chemistry A 123~(6) (2019) 1225--1232.

\bibitem{Kleist2019(2)}
E.~M. Kleist, T.~M. Korter, Quantitative analysis of minium and vermilion mixtures using low-frequency vibrational spectroscopy, Analytical chemistry 92~(1) (2019) 1211--1218.

\bibitem{Moffa2024}
C.~Moffa, C.~Merola, E.~Chiadroni, L.~Giuliani, A.~Curcio, L.~Palumbo, A.~C. Felici, M.~Petrarca, Pigments, minerals, and copper-corrosion products: Terahertz continuous wave (thz-cw) spectroscopic characterization of antlerite and atacamite, Journal of Cultural Heritage 66 (2024) 483--490.

\bibitem{Moffa2025}
C.~Moffa, A.~C. Felici, M.~Petrarca, Terahertz investigation of cultural heritage synthetic materials: A case study of copper silicate pigments, Minerals 15~(5) (2025) 490.

\bibitem{Moffa2025(2)}
C.~Moffa, D.~Francescone, A.~Curcio, A.~C. Felici, M.~Bellaveglia, L.~Piersanti, M.~Migliorati, M.~Petrarca, Deciphering hidden layers’s images through terahertz spectral fingerprints, Spectrochimica Acta Part A: Molecular and Biomolecular Spectroscopy (2025) 126510.

\bibitem{Fukunaga2023}
K.~Fukunaga, Nondestructive evaluation of lined paintings by thz pulsed time-domain imaging, Heritage 6~(4) (2023) 3448--3460.

\bibitem{Fukunaga2024}
K.~Fukunaga, Y.~Ueno, C.~Watanabe, A.~Yanagida, S.~Wakiya, Nondestructive observation of multilayered paintings on a single canvas by thz time-domain imaging and x-ray fluorescence elemental mapping, Journal of Infrared, Millimeter, and Terahertz Waves (2024) 1--17.

\bibitem{Moffa2024(2)}
C.~Moffa, A.~Curcio, C.~Merola, M.~Migliorati, L.~Palumbo, A.~C. Felici, M.~Petrarca, Discrimination of natural and synthetic forms of azurite: an innovative approach based on high-resolution terahertz continuous wave (thz-cw) spectroscopy for cultural heritage., Dyes and Pigments (2024) 112287.

\bibitem{Moffa2024(3)}
C.~Moffa, V.~Urso, M.~Migliorati, L.~Palumbo, A.~Felici, G.~Zollo, M.~Petrarca, Terahertz time-domain investigation of atacamite: Spectral analysis and theoretical insights for cultural heritage applications, Applied Physics Letters 125~(20) (2024).

\bibitem{Naftaly2015}
M.~Naftaly, Terahertz metrology, Artech House, 2015.

\bibitem{Bigourd2006}
D.~Bigourd, A.~Cuisset, F.~Hindle, S.~Matton, E.~Fertein, R.~Bocquet, G.~Mouret, Detection and quantification of multiple molecular species in mainstream cigarette smoke by continuous-wave terahertz spectroscopy, Optics letters 31~(15) (2006) 2356--2358.

\bibitem{Hindle2008}
F.~Hindle, A.~Cuisset, R.~Bocquet, G.~Mouret, Continuous-wave terahertz by photomixing: applications to gas phase pollutant detection and quantification, Comptes Rendus Physique 9~(2) (2008) 262--275.

\bibitem{Galstyan2021}
V.~Galstyan, A.~D’Arco, M.~Di~Fabrizio, N.~Poli, S.~Lupi, E.~Comini, Detection of volatile organic compounds: From chemical gas sensors to terahertz spectroscopy, Reviews in Analytical Chemistry 40~(1) (2021) 33--57.

\bibitem{DArco2022}
A.~D’Arco, D.~Rocco, F.~P. Magboo, C.~Moffa, G.~Della~Ventura, A.~Marcelli, L.~Palumbo, L.~Mattiello, S.~Lupi, M.~Petrarca, Terahertz continuous wave spectroscopy: A portable advanced method for atmospheric gas sensing, Optics Express 30~(11) (2022) 19005--19016.

\bibitem{Tyree2022}
D.~J. Tyree, P.~Huntington, J.~Holt, A.~L. Ross, R.~Schueler, D.~T. Petkie, S.~S. Kim, C.~C. Grigsby, C.~Neese, I.~R. Medvedev, Terahertz spectroscopic molecular sensor for rapid and highly specific quantitative analytical gas sensing, ACS sensors 7~(12) (2022) 3730--3740.

\bibitem{Wang2022}
W.~Wang, N.~Zhu, Z.~Wang, C.~Zhao, Z.~Song, X.~Chen, X.~Chao, Efficient terahertz absorption gas sensor with gaussian process regression in time-and frequency-domain, Sensors and Actuators B: Chemical 369 (2022) 132349.

\bibitem{Passarelli2022}
A.~Passarelli, T.~E. Rice, M.~A.~Z. Chowdhury, M.~N. Powers, M.~W. Mansha, I.~Wilke, M.~M. Hella, M.~A. Oehlschlaeger, Terahertz-wave absorption gas sensing for dimethyl sulfoxide, Applied Sciences 12~(11) (2022) 5729.

\bibitem{Powers2023}
M.~N. Powers, T.~E. Rice, A.~Chowdhury, M.~W. Mansha, M.~M. Hella, I.~Wilke, M.~A. Oehlschlaeger, Dimethyl ether gas sensing using rotational absorption spectroscopy in the thz frequency region from 220 to 330 ghz, Sensors and Actuators B: Chemical 384 (2023) 133635.

\bibitem{Su2012}
K.~Su, L.~Moeller, R.~B. Barat, J.~F. Federici, Experimental comparison of terahertz and infrared data signal attenuation in dust clouds, JOSA A 29~(11) (2012) 2360--2366.

\bibitem{Demers2017}
J.~R. Demers, F.~Garet, J.-L. Coutaz, Atmospheric water vapor absorption recorded ten meters above the ground with a drone mounted frequency domain thz spectrometer, IEEE sensors letters 1~(3) (2017) 1--3.

\bibitem{Rice2021}
T.~Rice, M.~Chowdhury, M.~Mansha, M.~Hella, I.~Wilke, M.~Oehlschlaeger, Halogenated hydrocarbon gas sensing by rotational absorption spectroscopy in the 220--330 ghz frequency range, Applied Physics B 127~(8) (2021) 123.

\bibitem{Demers2018}
J.~R. Demers, J.-L. Coutaz, F.~Garet, L.~P. Sadwick, T.~Yang, A uav-mounted thz spectrometer for real-time gas analysis, SPIE 10531, Terahertz, RF, Millimeter, and Submillimeter-Wave Technology and Applications XI, 105310K (23 February 2018) (2018).

\bibitem{Moffa2025_arxiv}
C.~Moffa, A.~Curcio, C.~Merola, V.~M. Orsini, D.~Francescone, F.~J.~P. Magboo, M.~Magi, M.~Coppola, L.~Giuliano, M.~Migliorati, G.~Zollo, M.~Reverberi, L.~Mattiello, M.~Petrarca, \href{https://arxiv.org/abs/2505.23956}{Terahertz prototype for air pollutants detection}, submitted to \emph{Next Research} (2025).
\newblock \href {http://arxiv.org/abs/2505.23956} {\path{arXiv:2505.23956}}.
\newline\urlprefix\url{https://arxiv.org/abs/2505.23956}

\bibitem{Kong2018}
D.-Y. Kong, X.-J. Wu, B.~Wang, Y.~Gao, J.~Dai, L.~Wang, C.-J. Ruan, J.-G. Miao, High resolution continuous wave terahertz spectroscopy on solid-state samples with coherent detection, Optics express 26~(14) (2018) 17964--17976.

\bibitem{Vogt2019}
D.~W. Vogt, M.~Erkintalo, R.~Leonhardt, Coherent continuous wave terahertz spectroscopy using hilbert transform, Journal of Infrared, Millimeter, and Terahertz Waves 40 (2019) 524--534.

\bibitem{Slocum2013}
D.~M. Slocum, E.~J. Slingerland, R.~H. Giles, T.~M. Goyette, Atmospheric absorption of terahertz radiation and water vapor continuum effects, Journal of Quantitative Spectroscopy and Radiative Transfer 127 (2013) 49--63.

\bibitem{Neumaier2015}
P.~Neumaier, K.~Schmalz, J.~Borngr{\"a}ber, D.~Kissinger, H.-W. H{\"u}bers, Terahertz gas-sensors: Gas-phase spectroscopy and multivariate analysis for medical and security applications, in: 2015 IEEE SENSORS, IEEE, 2015, pp. 1--4.

\bibitem{Hepp2016}
C.~Hepp, S.~L{\"u}ttjohann, A.~Roggenbuck, A.~Deninger, S.~Nellen, T.~G{\"o}bel, M.~J{\"o}rger, R.~Harig, A cw-terahertz gas analysis system with ppm detection limits, in: 2016 41st International Conference on Infrared, Millimeter, and Terahertz waves (IRMMW-THz), IEEE, 2016, pp. 1--2.

\bibitem{Neese2012}
C.~F. Neese, I.~R. Medvedev, G.~M. Plummer, A.~J. Frank, C.~D. Ball, F.~C. De~Lucia, Compact submillimeter/terahertz gas sensor with efficient gas collection, preconcentration, and ppt sensitivity, IEEE Sensors Journal 12~(8) (2012) 2565--2574.

\bibitem{Fosnight2013}
A.~M. Fosnight, B.~L. Moran, I.~R. Medvedev, Chemical analysis of exhaled human breath using a terahertz spectroscopic approach, Applied Physics Letters 103~(13) (2013).

\bibitem{Cheville1995}
R.~Cheville, D.~Grischkowsky, Far-infrared terahertz time-domain spectroscopy of flames, Optics letters 20~(15) (1995) 1646--1648.

\bibitem{Cheville1999}
R.~Cheville, D.~Grischkowsky, Far-infrared foreign and self-broadened rotational linewidths of high-temperature water vapor, JOSA B 16~(2) (1999) 317--322.

\bibitem{Matton2006}
S.~Matton, F.~Rohart, R.~Bocquet, G.~Mouret, D.~Bigourd, A.~Cuisset, F.~Hindle, Terahertz spectroscopy applied to the measurement of strengths and self-broadening coefficients for high-j lines of ocs, Journal of molecular spectroscopy 239~(2) (2006) 182--189.

\bibitem{Bigourd2007}
D.~Bigourd, A.~Cuisset, F.~Hindle, S.~Matton, R.~Bocquet, G.~Mouret, F.~Cazier, D.~Dewaele, H.~Nouali, Multiple component analysis of cigarette smoke using thz spectroscopy, comparison with standard chemical analytical methods, Applied Physics B 86 (2007) 579--586.

\bibitem{Cazzoli2008}
G.~Cazzoli, C.~Puzzarini, G.~Buffa, O.~Tarrini, Pressure-broadening of water lines in the thz frequency region: Improvements and confirmations for spectroscopic databases.: Part i, Journal of Quantitative Spectroscopy and Radiative Transfer 109~(17-18) (2008) 2820--2831.

\bibitem{Cazzoli2009}
G.~Cazzoli, C.~Puzzarini, G.~Buffa, O.~Tarrini, Pressure-broadening of water lines in the thz frequency region: Improvements and confirmations for spectroscopic databases. part ii, Journal of Quantitative Spectroscopy and Radiative Transfer 110~(9-10) (2009) 609--618.

\bibitem{Svelto2010}
O.~Svelto, D.~C. Hanna, et~al., Principles of lasers, Vol.~1, Springer, 2010.

\bibitem{Ren2010}
Y.~Ren, J.~Hovenier, R.~Higgins, J.~Gao, T.~Klapwijk, S.~Shi, A.~Bell, B.~Klein, B.~Williams, S.~Kumar, et~al., Terahertz heterodyne spectrometer using a quantum cascade laser, Applied Physics Letters 97~(16) (2010).

\end{thebibliography}

%\begin{thebibliography}{00}

%% For numbered reference style

\end{document}